\documentclass[aps,prx, reprint, superscriptaddress,nofootinbib]{revtex4-2}
\usepackage{graphicx}
\usepackage{dcolumn}
\usepackage{dsfont}
\usepackage{amssymb}
\usepackage{color}
\usepackage{bm}
\usepackage{amsmath}
\usepackage{gensymb}
\usepackage{float}
\usepackage{booktabs} 
\usepackage{tabularx}
\usepackage{mhchem}
\usepackage{hyperref}

\begin{document}

\title{Linear magneto-birefringence as a probe of altermagnetism}

\author{V. Sunko}
\email{vsunko@ista.ac.at}
\affiliation {Institute of Science and Technology Austria, Am Campus 1, 3400 Klosterneuburg, Austria }

\author{J. Orenstein}
\affiliation {Department of Physics, University of California, Berkeley, California 94720, USA}
\affiliation {Materials Science Division, Lawrence Berkeley National Laboratory, Berkeley, California 94720, USA}

\begin{abstract}
Altermagnets are a class of collinear magnets that exhibit non-relativistic spin splitting (NRSS) of electronic bands in the absence of net magnetization. Their potential to generate large spin polarization without spin-orbit coupling has created strong interest in probes that access the underlying order parameter directly. In this Perspective, we show that linear magneto-birefringence (LMB) provides a natural and broadly applicable route to detecting altermagnetic order. Building on the correspondence between the momentum-space structure of NRSS and the ferroic ordering of magnetic multipoles in real space, we demonstrate how $d$-wave and $g$-wave NRSS textures yield distinct LMB responses. We present a symmetry-based framework that identifies the optical geometries and field configurations required to isolate specific multipole components, enabling domain imaging and providing benchmarks for theoretical models of LMB.
\end{abstract}

\maketitle

\section{Introduction}

There is growing interest in materials that exhibit non-relativistic spin splitting (NRSS) of electronic bands without a net magnetization~\cite{hayami_momentum_2019,yuan_giant_2020, yuan_prediction_2021, smejkal_beyond_2022, smejkal_emerging_2022, bai_altermagnetism_2024, cheong_altermagnetism_2025}, called altermagnets. The promise of NRSS lies in its potential to generate large spin splittings, comparable to the electronic bandwidth, without relying on spin-orbit coupling (SOC) or ferromagnetic order. This makes altermagnets attractive for spintronics applications where large spin polarization and fast control are essential. 

Altermagnetism has been predicted in a variety of magnetic structures~\cite{guo_spin-split_2023, bai_altermagnetism_2024, wan_high-throughput_2025, gao_ai-accelerated_2025}, making unambiguous characterization of the symmetry of the ordered state challenging. The task is further complicated by the presence of domains related by time reversal symmetry (TRS) or point group symmetries~\cite{amin_nanoscale_2024, yamamoto_altermagnetic_2025}. As the desired properties of altermagnets, such as directionally dependent spin current, average out in multidomain samples, there is a premium on spatially resolved probes of altermagnetic domains.

Optical techniques are well-suited for detection of altermagnetic order and domain imaging as they are spatially-resolved probes of symmetry breaking. While the magneto-optical Kerr effect (MOKE) and circular dichroism (CD) have been used for detecting altermagnetism~\cite{hubert_anomalous_2025, amin_nanoscale_2024, yamamoto_altermagnetic_2025, hariki_x-ray_2024, galindez-ruales_revealing_2025}, this approach has clear limitations.  Firstly, MOKE and CD are not allowed in all altermagnets. Second, both effects require SOC, which is conceptually at odds with the defining feature of altermagnets: their ability to generate spin splitting without SOC.

Fortunately, MOKE and CD are not the only optical probes of TRS-breaking in antiferromagnets (AFMs). Since the 1970s, it has been understood from a symmetry perspective that certain AFMs may exhibit optical anisotropy that grows linearly with magnetic field (Refs.~\cite{eremenko_magneto-optics_1987, kharchenko_linear_1994}, and references within). This  \textit{linear magneto-birefringence} (LMB) effect has been detected in a number of materials which we now understand to be altermagnets, such as \ce{MnF2}~\cite{kharchenko_odd_2005, higuchi_control_2016}, \ce{CoF2}~\cite{kharchenko_lowering_1978}, \ce{CoCO3}~\cite{kharchenko_light_1978} and hematite~\cite{merkulov_linear_1981}. Recently, a general symmetry-based connection between altermagnetism and LMB has been pointed out~\cite{kimel_optical_2024}, LMB has been evaluated in a specific microscopic model~\cite{vila_orbital-spin_2025}, and related transport phenomena were discussed~\cite{zyuzin_linear_2021, vorobev_dwave_2024, sunko_linear_2025}.

The goal of this Perspective is to clarify the links between the family of LMB effects and altermagnetic order. As background we review the correspondence between the wavevector-dependence of the NRSS and the ferroic ordering of multipoles of real space magnetization density, originally discussed in Refs.~\cite{bhowal_ferroically_2024, mcclarty_landau_2024}. We then describe how the various multipoles can be detected through measurements of the birefringence induced by a magnetic field, or through the combined application of strain and magnetic field. Following this largely theoretical discussion, we conclude with a description of the experimental scheme for performing sensitive, spatially resolved measurements of LMB phenomena. 

\section{Magnetic multipoles and altermagnetism}

\begin{figure*}[t]
\centering
\includegraphics[width=\linewidth]{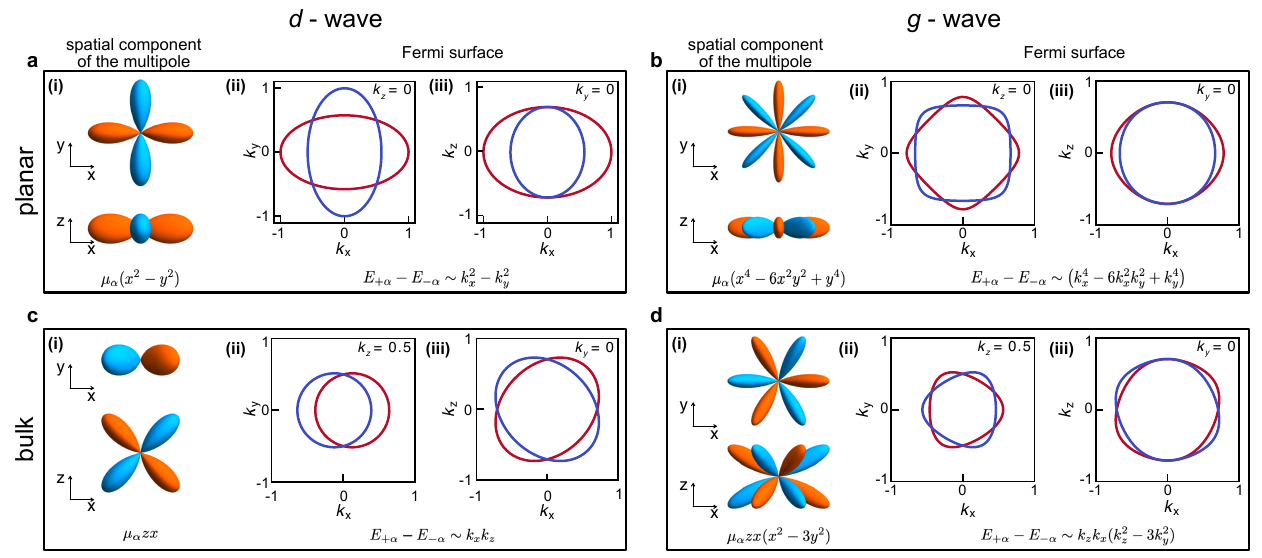}
\caption{\textbf{Momentum-space spin textures and corresponding real-space magnetic multipoles.}
Examples of non-relativistic spin-splitting (NRSS) patterns classified as (a, c) \textit{d}-wave and (b, d) \textit{g}-wave, in both planar and bulk forms.
Panels (a, b) show ``planar'' structures, where spin-splitting does not change sign with $k_z$, and panels (c, d) show ``bulk'' structures, where spin-splitting changes sign with $k_z$.
For each structure we show: (i) spatial components of the real-space magnetic multipole; (ii) spin-resolved Fermi surfaces in the $k_x$-$k_y$ plane; and (iii) spin-resolved Fermi surfaces in the $k_x$-$k_z$ plane. For each multipole, we list its real space representation and energy dispersion in $k$-space.}
\label{fig:fermi_multipoles}
\end{figure*}

The symmetry of non-relativistic spin splitting (NRSS) in momentum space is directly linked to the ferroic order of magnetic multipoles. NRSS textures can be expressed by magnetization density $\mu_l(\bm{k})$, where $l$ labels the magnetization direction and $\mu_l(\bm{k})$ is a polynomial in the momentum components. For example, a $d$-wave splitting has the form $\mu_l(\bm{k})\propto k_i k_j$, while a $g$-wave splitting follows $\mu_l(\bm{k}) \propto k_i k_j k_p k_q$. Each of these textures maps to a specific real-space multipole~\cite{bhowal_ferroically_2024}, whose ordering is basis for constructing Landau free energy functionals and predicting  altermagnetic observables~\cite{mcclarty_landau_2024}.

To illustrate the mapping, we first define real-space magnetic multipoles. The magnetization density $\mu(\bm{r})$ can be formally expressed as an expansion in powers of $r_i$ whose coefficients are given by,
\begin{equation}\label{eq:volume}
\mathcal{M}_{ijk\ldots l} = \int_{V} \mu_i(\bm{r})\, r_j r_k \ldots r_l \, d^3\bm{r},
\end{equation}
where $\mathcal{M}_{ijk\ldots l}$ is a multipole of rank equal to the number of indices $i,j,k,\ldots,l$ and $\mu_i(\bm{r})$ is the magnetization density along direction $i=x,y,z$. The spin-splitting in altermagnets is determined by the lowest order even-parity multipoles, which are the magnetic octupoles (third rank: two powers of $\mathbf{r}$ and one of $\bm{\mu}$) and triakontadipoles (fifth rank: four powers of $\mathbf{r}$ and one of $\bm{\mu}$).
As shown in Refs.~\cite{watanabe_group_2018,bhowal_ferroically_2024}, the mapping between momentum-space textures and real-space multipoles is straightforward, as long as multipoles are parity-even: replace factors of $\bm{k}$ with $\bm{r}$. Thus, a $d$-wave NRSS texture $\mu_l(\bm{k}) \propto k_i k_j$ corresponds to the octupole $\mu_l r_i r_j$~\cite{bhowal_ferroically_2024}, while a $g$-wave pattern maps to a triakontadipole $\mu_l r_i r_j r_p r_q$~\cite{verbeek_nonrelativistic_2024}. 

The four panels of Fig.~\ref{fig:fermi_multipoles} illustrate this mapping for $d$-wave and $g$-wave altermagnetic order. For each, we show a planar and a bulk example~\cite{smejkal_beyond_2022}, where ``planar'' indicates that the splitting is even in $k_z$, and ``bulk'' indicates that it is odd. For example, sub-panels a(ii) and a(iii) of Fig.~\ref{fig:fermi_multipoles} depict the spin-up and spin-down Fermi surfaces (in red and blue, respectively) for a $d$-wave altermagnet whose NRSS is described by the dispersion,
\[E_\pm(k_x, k_y) = a k^2 \pm \delta(k_x^2 - k_y^2),
\]
where $k^2 = k_x^2 + k_y^2$, and $\pm$ denotes the two spin branches. The spatial part of the associated real-space multipole is obtained by replacing $k_x$ and $k_y$ with $x$ and $y$, yielding $(x^2 - y^2)$, shown in Fig.~\ref{fig:fermi_multipoles}a(i).

Note that we have not specified a spin quantization axis, reflecting the absence of spin-orbit coupling (SOC) in the nonrelativistic limit. In real materials, SOC is always present and selects a preferred spin axis. For clarity, we adopt an informal but intuitive notation in which we write multipole moments in the form, $\mathcal{M}_{\alpha,\beta}$, where $\alpha = x, y, z$ specifies the spin quantization axis and $\beta$ is a form factor. For example, an octupolar moment with $\beta=x^2-y^2$ is given by, 
\begin{equation}\label{eq:x2y2}
\mathcal{M}_{\alpha,x^2 - y^2} = \int_V \mu_\alpha(\bm{r})(x^2 - y^2) \, d^3\bm{r}.
\end{equation}
For a more formal notation and tensor decomposition of magnetic octupoles we refer the reader to Ref.~\cite{urru_magnetic_2022}, and references therein.

\section{Optical probes of multipolar order}

Having introduced the connection between altermagnetic order and magnetic multipoles, we turn to the detection of multipolar order using optical probes. At our disposal are several experimental control parameters: the electric field of the light wave ($E_i$), an external magnetic field ($H_i$), and uniaxial strain ($\eta_{ij}$, instead of more common $\epsilon_{ij}$ to avoid confusion with the dielectric tensor). The key is to identify which combinations of these fields share the symmetry of the multipolar order parameters of interest. 

To that end, we note that $r_i$ transforms like the electric field $E_i$; the product $r_i r_j$ transforms like the strain tensor $\eta_{ij}$; and the magnetic field $H_i$ transforms like the magnetic moment $\mu_i$. These correspondences allow us to systematically construct combinations of experimental fields that transform under the same irreducible representations as the relevant magnetic multipoles (see Table~\ref{tab:MultipoleCoupling}). Because the free energy must remain invariant under all symmetry operations of the parent (paramagnetic) space group, it can only contain terms in which the order parameter couples to combinations of external fields with matching symmetry. For example, in the case of a $d$-wave altermagnet the free energy, $F_d$, has terms,
\begin{equation}\label{eq:FreeD}
F_{d} =F_{d}^0+ a\, \mathcal{M}_{ijk} E_j E_k H_i + b\, \mathcal{M}_{ijk} \eta_{jk} H_i,
\end{equation}

\noindent where $F_{d}^0$ is the part of the free energy that is independent of $\mathcal{M}_{ijk}$. For a $g$-wave altermagnet we have:
\begin{align}\label{eq:FreeG}
F_g = F_g^0 &+ a\, \mathcal{M}_{ijklm} E_j E_k E_l E_m H_i 
\notag \\
&+ b\,  \mathcal{M}_{ijklm} E_j E_k \eta_{lm}H_i + c\,  \mathcal{M}_{ijklm} \eta_{jk} \eta_{lm} H_i,
\end{align}
\noindent with $F_{g}^0$ denoting part of the free energy that is independent of $\mathcal{M}_{ijklm}$. 

This construction of the free energy directly identifies experimental probes are sensitive to each multipole, providing a framework for their detection. For example, the piezomagnetic effect, whereby an applied strain induces a magnetization and associated phenomena, has been discussed in the context of $d$-wave magnetism and magnetic octupoles~\cite{bhowal_ferroically_2024, naka_nonrelativistic_2025, lei_shear-strain-induced_2025, chakraborty_magnetic_2025, sun_symmetry-breaking_2025, khodas_tuning_2025}. Inspection of Eq.~\ref{eq:FreeD} makes this association transparent: 
\begin{equation}
    M_i\equiv\frac{\partial F_d}{\partial H_i}\propto \mathcal{M}_{ijk}\eta_{jk}
\end{equation}
where $M_i$ is the $i$-component of the magnetization.

\begin{table}[t]
\centering
\begin{tabular}{|l|l|l|}
\hline
\textbf{NRSS} & \textbf{Multipole} & \textbf{Coupling fields} \\
\hline
$s$-wave & $\mu_i$ & $H_i$ \\ \hline
$d$-wave & $\mu_i r_j r_k$ & $E_j E_k H_i$, $\eta_{jk} H_i$ \\  \hline
$g$-wave & $\mu_i r_j r_k r_l r_m$ &
$E_j E_k E_l E_m H_i$, $\eta_{jk}\eta_{lm}H_i$, $\eta_{jk}E_l E_m H_i$ \\  
\hline
\end{tabular}
\caption{Even-parity magnetic multipoles, associated non-relativistic spin-splitting (NRSS) patterns, and symmetry-allowed combinations of external fields that couple to them in the free energy. $s$-wave pattern corresponds to ferromagnetic order. The multipoles are found by integrating the quantities in the table over the volume, as defined in Eq.~\ref{eq:volume}.} 
\label{tab:MultipoleCoupling}
\end{table}

Here, we focus on the optical response embodied in the dielectric function, $\epsilon(\omega)$. While we cannot compute the frequency dependence response from the free energy, the symmetry constraints are independent of frequency: if a tensor component is allowed at $\omega=0$, it is allowed at $\omega\neq0$. The static dielectric tensor is given in terms of the free energy by,  
\begin{equation}
\frac{\epsilon_{jk}}{\epsilon_0}-1=  \chi_{jk}=\frac{\partial P_j}{\partial E_k}=\frac{\partial^2 F}{\partial E_k\partial E_j},
\end{equation}
where $P_j$ is the $j$-th component of polarization,  $\chi_{ij}$ the electronic susceptibility and $\epsilon_0$ is the vacuum permittivity. The coupling of the dielectric tensor to the multipolar order appears in the terms in Eq.~\ref{eq:FreeD} and Eq.~\ref{eq:FreeG} that are second order in the electric field. Below we provide some specific examples showing how the symmetry of multipolar order can be identified by measurements of the dielectric tensor.

\subsection{Octupoles: Linear Magneto-Birefringence}

Based on the free energy associated with octupole order (Eq.~\ref{eq:FreeD}), we find a contribution to the dielectric tensor of the form
\begin{equation} \label{eq:LMB}
\epsilon_{jk} = \epsilon_{jk}^0 + a\, \mathcal{M}_{ijk} H_i.
\end{equation}
As the multipole moments are symmetric under permutation of the spatial indices $j,k,...l$ (see Eq.~\ref{eq:volume}), they contribute only to the symmetric part of the dielectric tensor ($\epsilon_{jk}=\epsilon_{kj}$). As we discuss more fully in Section IV, the symmetric part of $\epsilon_{jk}$ manifests as optical birefringence. The contribution to the symmetric part that is linear in field is what we have referred to as the linear magneto-birefringence or LMB.  

The LMB effect, like MOKE and RCD, is an unambiguous indicator of broken time-reversal symmetry. This follows from Onsager's reciprocity relation, which stipulates that if a system is time-reversal-invariant then $\epsilon_{jk}(H) = \epsilon_{kj}(-H)$. Applying this condition to Eq.~\ref{eq:LMB} gives,
\begin{equation}
\epsilon^0_{jk}+\mathcal{M}_{ijk}H_i=\epsilon^0_{kj}-\mathcal{M}_{ikj}H_i=\epsilon^0_{jk}-\mathcal{M}_{ijk}H_i,
\end{equation}
where the last equality follows because  $\mathcal{M}_{ijk}=\mathcal{M}_{ikj}$ by definition. Therefore, time-reversal symmetry forbids a linear dependence of the symmetric part of the dielectric function on magnetic field, in other words the LMB effect.

In the following, we show, through a few examples, how the symmetry of the multipolar order can be distinguished by its contribution to the dielectric tensor. We leave the field orientation index $\alpha$ unspecified for now; it corresponds to the direction of magnetic moments in the material. While not essential for the conceptual discussion here, it is important for the design of experiments, where a direction of the magnetic field needs to be chosen.
\subsection*{Example 1: $\mathcal{M}_{\alpha, x^2 - y^2}$ Multipole}
As a first example, consider the $\mathcal{M}_{\alpha, x^2 - y^2}$ octupole (Eq.~\ref{eq:x2y2}). The corresponding free energy is,
\begin{equation}\label{eq:Freex2y2}
F_{d1} = F_{d1}^0 + a\, \mathcal{M}_{\alpha, x^2 - y^2} (E_x^2 - E_y^2) H_\alpha,
\end{equation}
which leads to the dielectric tensor,
\begin{equation}
\epsilon_{xx}(H)=-\epsilon_{yy}(H) = \epsilon^0 + a\, \mathcal{M}_{\alpha, x^2 - y^2} H_\alpha.
\end{equation}
In the presence of a magnetic field, the $\mathcal{M}_{\alpha, x^2 - y^2}$ octupole enhances the anisotropy between the diagonal components of LMB, $\epsilon_{xx}-\epsilon_{yy}$; we call this effect the diagonal LMB.

\subsection*{Example 2: $\mathcal{M}_{\alpha, xy}$ Multipole}
Next, we consider the $\mathcal{M}_{\alpha, xy}$ octupole, which can be obtained from the $\mathcal{M}_{\alpha, x^2 - y^2}$ octupole by a rotation of $45\degree$ around the $z$ axis, and is found in \ce{MnF2}~\cite{bhowal_ferroically_2024}. Applying the same approach as above, we find:
\begin{equation}\label{eq:FreeXY}
F_{d2} = F_{d2}^0 + a\, \mathcal{M}_{\alpha, xy} E_x E_y H_\alpha,
\end{equation}
which gives:
\begin{equation}
\epsilon_{xy}(H)=\epsilon_{yx}(H) = \epsilon^0 + a\, \mathcal{M}_{\alpha, xy} H_\alpha.
\end{equation}
In this case the magnetic field enhances the off-diagonal anisotropy; we call it the off-diagonal LMB. 

\subsection*{Example 3: $\mathcal{M}_{\alpha, zx}$ Multipole}

Finally, consider the $\mathcal{M}_{\alpha, zx}$ octupole, which yields the following contribution to the dielectric tensor,
\begin{equation}
\epsilon_{xz}(H)=\epsilon_{zx}(H) = \epsilon^0 + a\, \mathcal{M}_{\alpha, zx} H_\alpha.
\end{equation}
Conceptually, this case is similar to the previous examples. However, since $\mathcal{M}_{\alpha, zx}$ couples to $\epsilon_{xz}$, it cannot be detected by a normally incident wave propagating along $z$. This limitation can be overcome in one of two ways: either by performing the experiment at oblique incidence, or by reorienting the sample to expose the $zx$ face. This example highlights the importance of careful experimental geometry design in detecting magnetic multipoles.

\subsection{Triakontadipoles: Strain Linear Magneto-Birefringence}

Triakontadipoles in $g$-wave altermagnets give rise to higher-order terms in the free energy (Eq.~\ref{eq:FreeG}), which in turn generate higher-order contributions to the dielectric tensor,
\begin{equation} \label{eq:LMB_g}
\epsilon_{jk} = \epsilon_{jk}^0 + a\, \mathcal{M}_{ijklm} E_l E_m H_i + b\, \mathcal{M}_{ijklm} \eta_{lm} H_i.
\end{equation}
In the absence of strain, the lowest-order optical effect arising  from the multipole is a dielectric tensor term proportional to the magnetic field and two powers of the electric field.  It could be detected through third-harmonic generation under an applied magnetic field. To remain within the domain of linear optics, we focus on the second term in Eq.~\ref{eq:LMB_g}, which shows that $g$-wave altermagnets produce a response proportional to both strain and magnetic field. Unlike electric fields of the light wave, strain cannot be freely rotated in experiments: each strain orientation typically requires a separately prepared sample. The form of the multipole determines which strain orientations should be used. We illustrate this with two examples.

\subsection*{Example 1: $\mathcal{M}_{\alpha, x^4 - 6x^2y^2 + y^4}$ Multipole}

First we consider the planar $g$-wave multipole shown in Fig.~\ref{fig:fermi_multipoles}b. By inspection, we find the combination of strain and fields that couples linearly to the multipole:
\begin{align}\label{eq:g1}
F_{g1} = &F_{g1}^0 + b\, \mathcal{M}_{\alpha, x^4 - 6x^2y^2 + y^4} \times \notag \\
&\left[ (\eta_{xx} - \eta_{yy})(E_x^2 - E_y^2) - 4\eta_{xy} E_x E_y \right] H_\alpha.
\end{align}
By comparing Eq.~\ref{eq:g1} with the free energy of $d$-wave altermagnets, Eq.~\ref{eq:Freex2y2} and Eq.~\ref{eq:FreeXY}, we see that $(\eta_{xx} - \eta_{yy})$ strain induces diagonal LMB, while $\eta_{xy}$ induces off-diagonal LMB.

\subsection*{Example 2: $\mathcal{M}_{\alpha, xz(x^2 - 3y^2)}$ Multipole}

Next, we consider the bulk $g$-wave multipole shown in Fig.~\ref{fig:fermi_multipoles}d. Two strain components that couple to $\mathcal{M}_{\alpha, xz(x^2 - 3y^2)}$ are $\eta_{xz}$ and $(\eta_{xx} - \eta_{yy})$. The corresponding free energy is:
\begin{align}\label{eq:g2}
&F_{g2} = F_{g2}^0 + b\, \mathcal{M}_{\alpha, xz(x^2 - 3y^2)} \left[ 2\eta_{xz} \left( E_x^2 - E_y^2 - \frac{E^2}{2} \right) \right] H_\alpha \notag \\
&+ b'\, \mathcal{M}_{\alpha, xz(x^2 - 3y^2)} \left[ 2E_x E_z \left( \eta_{xx} - \eta_{yy} - \frac{\eta_{xx} + \eta_{yy}}{2} \right) \right] H_\alpha,
\end{align}
where $E^2 = E_x^2 + E_y^2$. At normal incidence, with fields in the $xy$ plane, diagonal LMB is induced only by $\eta_{xz}$, while $\eta_{xx} - \eta_{yy}$ induces off-diagonal LMB in the $xz$ plane.  Direct detection of the bulk $g$-wave response requires engaging with all three spatial directions, $x$, $y$, and $z$, either through strain or electric fields in the light wave. This makes experimental detection more challenging than in the planar examples. 

We note that if we were to rotate $\mathcal{M}_{\alpha, xz(x^2 - 3y^2)}$ by $90\degree$ about the $x$ axis, we obtain the $\mathcal{M}_{\alpha, yz(y^2 - 3x^2)}$, which is active in, for example, MnTe~\cite{mcclarty_landau_2024}. The corresponding free energy can be found by replacing $x\rightarrow y$ and $y\rightarrow -x$ in Eq.~\ref{eq:g2}. It has been recognized that strained MnTe gives rise to a large $d$-wave spin-splitting~\cite{belashchenko_giant_2025}, consistent with our analysis.

\section{Measurement of linear magneto-birefringence}

\begin{figure}
    \centering
    \includegraphics[width=\linewidth]{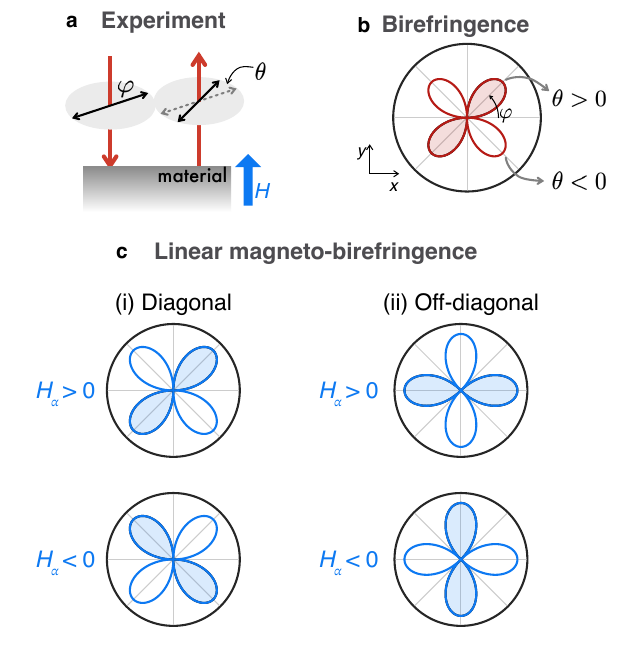}
   \caption{(a) Reflection geometry for measuring birefringence and linear magneto-birefringence (LMB) at normal incidence: linearly polarized light, with polarization angle $\varphi$, is focused onto the material; upon reflection, the polarization is rotated by an angle $\theta$. (b) Birefringence signal: polar plot of $\theta$ vs.\ $\varphi$, where nodal directions indicate the principal optical axes ($\phi_0=0$). (c) Field-linear modification of birefringence: (i) diagonal LMB with unchanged axis orientation ($\phi_H=\phi_0$); (ii) off-diagonal LMB with axis rotation by $\pi/4$ ($\phi_H=\phi_0+\pi/4$).}  \label{fig:LMB_Intro}
\end{figure}

We focus on measurements of the linear optical response at normal-incidence. The observables are the reflection and transmission matrices ($\bm{R}$, $ \bm{T}$), which can be obtained from $\bm{\epsilon}(\omega,\bm{q}=0)$ via the Fresnel equations. The same symmetry constraints that apply to $\bm{\epsilon}$ appear in the matrices $\bm{R}$ and $\bm{T}$ as well. Since NRSS is especially relevant in metals, we focus here on the reflection geometry depicted in Fig.~\ref{fig:LMB_Intro}a, but the conclusions are valid for both experiments. 

The symmetric part of the reflectivity matrix at zero applied field can be parametrized as,
\begin{equation}\label{eq:RDecomp}
    \bm{R}^s = r^s_0 \mathds{1}
    + \delta r^s \begin{pmatrix}
        \cos(2\phi_0) & \sin(2\phi_0) \\
        \sin(2\phi_0) & -\cos(2\phi_0)
    \end{pmatrix},
\end{equation}

\noindent where $\delta r^s$ is the magnitude of the optical anisotropy and $\phi_0$ denotes the orientation of principal optical axes. The state of polarization of incident light in the laboratory frame is given by $(\cos{(\varphi)}, \sin{(\varphi)})$, with $\varphi$ the polarization angle. Both $r^s_0$ and $\delta r^s$ may be complex and while different measurement schemes probe $\operatorname{Re}(\delta r^s/r^s_0)$ and $\operatorname{Im}(\delta r^s/r^s_0)$, the same symmetry principles apply to both. For illustrative purposes we focus on $\operatorname{Re}(\delta r^s/r^s_0)$, which determines the rotation of linear polarization upon reflection from a birefringent medium. The rotation angle, $\theta$, depends on $\varphi$ and vanishes when $\varphi$ aligns with the principal axes, i.e. at $\varphi=\phi_0$ and $\varphi=\phi_0+\pi/2$.  Measurements of $\theta$ that reveal both the magnitude and orientation of the birefringence can be performed using a balanced detection scheme as described in e.g. Ref.~\cite{sunko_spin-carrier_2023}.  Fig.~\ref{fig:LMB_Intro}b shows a polar plot of the balanced detector output (proportional to $\theta$) as a function of $\varphi$, where the nodes correspond to the optical axes.

A nonzero LMB effect appears as a modulation of the reflectivity matrix that is linear in applied magnetic field,

\begin{equation}\label{eq:RDecomp2}
    \bm{R}^s(H_\alpha) = \bm{R}^s(0)+ \beta_\alpha H_\alpha \begin{pmatrix}
        \cos(2\phi_H) & \sin(2\phi_H) \\
        \sin(2\phi_H) & -\cos(2\phi_H)
    \end{pmatrix},
\end{equation}

\noindent where $H_\alpha$ is the field along direction $\alpha$ and $\beta_\alpha$ is a material-specific coefficient.  Polar plots of the component of the detector output that is linear in $H_\alpha$ are shown in Fig.~\ref{fig:LMB_Intro}c. Panel \ref{fig:LMB_Intro}c(i) illustrates diagonal LMB, where the field modulates the amplitude of the birefringence while the orientation of the fast and slow axes remains the same as in zero-field ($\phi_H=\phi_0$). Referring to the Example 1 considered in Section III.A, this observation would indicate $\mathcal{M}_{\alpha,x^2-y^2}$ multipolar order.  Alternatively, for off-diagonal birefringence ($\phi_H=\phi_0+\pi/4$) the field rotates the principal optical axes (Fig.~\ref{fig:LMB_Intro}c(ii)), consistent with the $\mathcal{M}_{\alpha,xy}$ multipolar order considered in Example 2 in Section III.A. LMB can be detected either by measuring linear dichroism as a function of static magnetic field~
\cite{kharchenko_odd_2005, higuchi_control_2016,kharchenko_lowering_1978,kharchenko_light_1978,merkulov_linear_1981}, or by detecting the changes in linear dichroism that are synchronous with an oscillating magnetic field~\cite{sunko_spin-carrier_2023, donoway_multimodal_2024}.

Finally, we note that in altermagnets, LMB can appear in the absence  of zero-field birefringence. In that case panels in Fig.~\ref{fig:LMB_Intro}c illustrate the total birefringence in an applied magnetic field. 

\section{Outlook}
We have outlined a correspondence between NRSS and LMB: each NRSS pattern is associated with a specific geometry of LMB. However, the converse is not true: observation of LMB does not uniquely point to a NRSS pattern, as the effect could also arise due to \textit{relativistic} spin-orbit coupling (SOC). Let us consider a concrete example. MnTe is a $g$-wave altermagnet, so our analysis shows that strain induces LMB. However, already the magnetic point group of \textit{unstrained} MnTe, $m'm'm$, allows for LMB; indeed, the piezomagnetic effect, which shares the symmetry of LMB, has been observed~\cite{aoyama_piezomagnetic_2024}. The apparent discrepancy arises because of SOC: in the absence of SOC, $g$-wave NRSS would generate LMB only in strained MnTe, but SOC enables additional effects, including LMB in the unstrained material. 

While this ambiguity may complicate the interpretation of LMB experiments, it also points to a path toward realizing the full potential of LMB as a probe of altermagnetism. Disentangling the roles of non-relativistic and relativistic mechanisms requires a combined theoretical and experimental effort that goes beyond purely symmetry-based reasoning, since symmetry alone cannot reveal the microscopic origin of LMB. We propose that progress can be achieved through joint theory-experiment studies of the LMB spectrum. By measuring and comparing LMB spectra across materials with distinct multipolar patterns and comparing them to first-principles calculations, spectral signatures unique to NRSS could be established. Our analysis provides the foundation for this effort: in $d$-wave altermagnets, LMB can arise without SOC, whereas in $g$-wave cases, it depends on SOC. 

Even before a microscopic understanding is established, LMB already serves as a symmetry-guided and experimentally accessible probe of altermagnetism, with immediate utility for domain imaging and manipulation.

\section{Acknowledgements}
We thank Nicola Spaldin for valuable discussions. J.O. received support from the Quantum Materials (KC2202) program under the U.S. Department of Energy, Office of Science, Office of Basic Energy Sciences, Materials Sciences and Engineering Division under Contract No. DE-AC02-05CH11231, and the Gordon and Betty Moore Foundation's EPiQS Initiative through Grant GBMF4537 to J.O. at UC Berkeley.

\bibliography{References}

\end{document}